\documentstyle[12pt,epsf]{article}
\begin{document}
\begin{titlepage}
\begin{flushright}
{\setlength{\baselineskip}{0.18in}
{\normalsize
METU-PHYS-HEP97/20\\
}}
\end{flushright}
\vskip 2cm
\begin{center}
{\Large\bf
Higgs-strahlung in Abelian Extended Supersymmetric Standard Model}

\vskip 1cm
           
{\large   
D. A. Demir\footnote{e-mail:ddemir@heraklit.physics.metu.edu.tr},
N. K. Pak\footnote{e-mail:npak@heraklit.physics.metu.edu.tr} \\}

\vskip 0.5cm
{\setlength{\baselineskip}{0.18in}
{\normalsize\it Middle East Technical University, Department of Physics, 
06531, Ankara, Turkey\\} }
\end{center}
\vskip .5cm

\begin{abstract}
We analyze the Higgs production via the Higgs-strahlung process 
$e^{+}e^{-}\rightarrow h_{k}{\it{l}}^{+}{\it{l}}^{-}$ in an Abelian 
Extended Supersymmetric SM. We work in the large Higgs trilinear coupling 
driven minimum of the potential, and find that the next-to-lightest Higgs 
cannot be produced by this process. Other Higgs scalars, namely the 
lightest and the heaviest, have cross sections comparable to that in 
the pure SM. It is found that the present model has observable 
differences with the other popular model, NMSSM, in the same type of 
minimum. 
\end{abstract}
\end{titlepage}
\newpage

\section{Introduction}
The main mechanism for Higgs production in $e^{+}e^{-}$ collisions at LEP2 
energies is the Higgs-strahlung process $e^{+}e^{-} \rightarrow Z H$ in 
which $H$ is radiated from a virtual $Z$ line. In the literature, this 
process has already been analyzed in SM \cite{sm-strah}, MSSM 
\cite{mssm-strah} and NMSSM \cite{nmssm-strah,lep2}. Hence, it would be 
convenient to discuss this collision process in other models, such as $Z'$ 
models, to confront the experimental results with all possible theoretical 
predictions. The $Z'$ models in which the SM gauge group is extended by 
an extra $U(1)$ follow directly from SUSY-GUT's or String 
compactifications. In addition this sound theoretical foundation, they 
have proven to be phenomenologically viable as well. For instance, like 
NMSSM, the $Z'$ models can solve the MSSM $\mu$ problem when SUSY is broken 
around the weak scale. However, they differ from MSSM and NMSSM in that
they are further constrained the phenomenologically required smallness 
of the $Z-Z'$ mixing angle. In \cite{lang}, a general string-inspired 
model was analyzed and phenomenologically viable portions of the 
parameter space have been identified. In this work we shall be working 
in one possible minimum constructed in \cite{lang}, that is, we shall 
consider that portion of the parameter space where the Higgs soft 
trilinear coupling is large compared to other soft masses in the Higgs 
sector. 
   
In a recent work \cite{biz} we have analyzed another important LEP2 
processs $e^{+}e^{-}\rightarrow (h A)\rightarrow b\bar{b}b\bar{b}$ in 
the Abelian Extended Supersymmetric Standard Model which was already 
introduced in \cite{lang}. In \cite{biz} we found that it was 
essentially the second neutral Higgs scalar $h_{2}$ contributing 
to the process in the limit of small $Z-Z'$ mixing angle. Now, using   
the model in \cite{biz}, we shall analyze the Higgs-strahlung type process
$e^{+}e^{-}\rightarrow h_{k}{\it{l}}^{+}{\it{l}}^{-}$, where
${\it{l}}=(e,\mu)$ and $h_{k}$ (k=1,2,3) be the CP-even Higgs scalars
of the model. As is well known from MSSM \cite{mssm-strah} Higgs-strahlung 
and pair production processes are complementary to each other and we 
expect similar behaviour to occur in the present model as well.

In Sec. 2 we shall reproduce the formulae relevant to the aim of the 
work, at the tree level. In Sec.3 we shall first analyze the cross section 
numerically, and discuss the results. Next, we consider the contribution 
of the radiative corrections at the one-loop order, and show that they do 
not alter the tree-level results significantly.

\section{Formalism}
The coupling of CP-even Higgs scalars $h_{k}$ (k=1,2,3) to vector bosons 
$Z_{i}$ (i=1,2) can be expressed as 
\begin{eqnarray}
{\it{L}}=K_{k i j} g_{\mu\nu}Z_{i}^{\mu}Z_{j}^{\nu}h_{k},
\end{eqnarray}
where the couplings $K_{ijk}$ are model dependent \cite{epsb}. In giving 
the expressions for $K_{ijk}$ in the present model it is convenient to 
work with two dimensionless quantities,that is, Higgs Yukawa coupling 
$h_{S}$, and normalized $U(1)_{Y'}$ coupling constant $\rho_{S}$ defined by
\begin{eqnarray} 
\rho_{S}=\frac{g_{Y'}Q_{S}}{G}
\end{eqnarray}
where $g_{Y'}$ and $Q_{S}$ are the $U(1)_{Y'}$ coupling constant and
$U(1)_{Y'}$ charge of the SM singlet, respectively. 

We shall first present the expressions for all the masses and couplings 
using the tree-level potential, deferring the discussion of effects of 
the radiative corrections to the next section. In the large 
trilinear coupling minimum \cite{lang}, CP-even Higgs masses have,  
in the increasing order, the following expressions: 
\begin{eqnarray} 
m_{h_{1}}\sim\frac{\sqrt{2}h_{S}M_{Z}}{G}\; , \; 
m_{h_{2}}\sim\sqrt{1+2\frac{h_{S}^{2}}{G^2}}M_{Z}\; , \; 
m_{h_{3}}\sim\sqrt{3\rho_{S}+2\frac{h^{2}_{S}}{G^2}}M_{Z},
\end{eqnarray}
where $G=\sqrt{g_{2}^{2}+g_{Y}^{2}}$, and $g_{2}$ and $g_{Y}$ are $SU(2)$ 
and $U(1)_{Y}$ coupling constants. Moreover, neutral gauge 
bosons are almost diagonal with the masses $M_{Z_{1}}=M_{Z}$, and 
$M_{Z_{2}}\sim\sqrt{3}\rho_{S}M_{Z}$. It is clear that $\rho_{S}$ measures 
the $Z'$ mass in units of $M_{Z}$ up to a factor of $\sqrt{3}$. Using 
the elements of the diagonalizing matrix $R$ defined in 
\cite{biz}, one can easily obtain the coupling constants $K_{kij}$ in (1) 
as listed below: 
\begin{eqnarray}
K_{111}&=&\frac{G 
M_{Z}}{\sqrt{6}}(\cos^{2}\alpha+3\rho_{S}^{2}\sin^{2}\alpha)\nonumber\\
K_{211}&=&\frac{G M_{Z}}{2}\rho_{S}\sin2\alpha\nonumber\\
K_{311}&=&-\frac{G 
M_{Z}}{\sqrt{12}}(\cos^{2}\alpha-3\rho_{S}^{2}\sin^{2}\alpha) \nonumber\\
K_{122}&=&\frac{G 
M_{Z}}{\sqrt{6}}(\sin^{2}\alpha+3\rho_{S}^{2}\cos^{2}\alpha)\nonumber\\
K_{222}&=&-\frac{G M_{Z}}{2}\rho_{S}\sin2\alpha\\
K_{322}&=&-\frac{G 
M_{Z}}{\sqrt{12}}(\sin^{2}\alpha-3\rho_{S}^{2}\cos^{2}\alpha) \nonumber\\
K_{112}&=&-\frac{2 G M_{Z}}{\sqrt{6}}\rho_{S}^{2}\sin2\alpha\nonumber\\
K_{212}&=&-G M_{Z}\rho_{S}\cos2\alpha\nonumber\\
K_{312}&=&-\frac{4 G M_{Z}}{\sqrt{12}}\rho_{S}^{2}\sin2\alpha\nonumber
\end{eqnarray}
where $\alpha$ is the $Z-Z'$ mixing angle, and is well 
below the phenomenological upper bound in the present minimum of the 
potential. The couplings $K_{kij}$ listed in (4) are interesting enough 
to have a detailed discussion of them here. If $\rho_{S}\sim 
{\cal{O}}(1)$ (which should be the case in phenomenologically acceptable 
$Z'$ models \cite{haber-sher}), one can safely neglect $\sin\alpha$ 
dependent terms all together. In this case the couplings $K_{ijk}$ get 
simplified and possess the following properties:
\begin{itemize}
\item $h_{1}$ and $h_{3}$ do have only $Z_{i}Z_{i}$ type couplings, that 
is, they have diagonal couplings in $(Z_{1}, Z_{2})$ space. 
\item Coupling costant of $h_{3}$ is always $-1/\sqrt{2}$ times the coupling 
constant of $h_{1}$, which is an immediate consequence of the 
diagonalizing matrix $R$ given in \cite{biz}. In this sense $h_{3}$ is a 
replica of $h_{1}$ with a larger mass. 
\item Coupling of $h_{2}$ requires always a transition from $Z_{1}$ to 
$Z_{2}$ and vice versa , that is, it does have only off-diagonal 
couplings in $(Z_{1}, Z_{2})$ space. $Z_{1}$, for example, yields 
$Z_{2}$ after radiating $h_{2}$; however, it yields $Z_{1}$ in the final 
state, after radiating $h_{1,3}$. After discussing the coupling of Higss 
scalars to fermions, the status of $h_{2}$ will be clearer. 
\end{itemize}

The vector coupling $v_{f}^{(i)}$ and axial coupling $a_{f}^{(i)}$ of Higgs 
scalars to the fermion $f$ and vector boson $Z_{i}$ are given by 
\begin{eqnarray}
\left(\begin{array}{c c} v^{(1)}_{f}\\v^{(2)}_{f}\end{array}\right)
&=&\left (\begin{array}{c c}
\cos\alpha&\sin\alpha\\\sin\alpha&-\cos\alpha\end{array}\right)
\left(\begin{array}{c c} v_{f}\\v'_{f}\end{array}\right)\\
\left(\begin{array}{c c} a^{(1)}_{f}\\a^{(2)}_{f}\end{array}\right)
&=&\left (\begin{array}{c c}
\cos\alpha&\sin\alpha\\\sin\alpha&-\cos\alpha\end{array}\right)
\left(\begin{array}{c c} a_{f}\\a'_{f}\end{array}\right)
\end{eqnarray}
where
\begin{eqnarray}
v_{f}&=&\frac{g_{2}}{4\cos\theta_{W}}(1-4\sin^{2}\theta_{W})\\
a_{f}&=&\frac{g_{2}}{4\cos\theta_{W}}\\
v'_{f}&=&\frac{g_{Y'}}{2}(Q_{fL}+Q_{fE})\\
a'_{f}&=&\frac{g_{Y'}}{2}(Q_{fL}-Q_{fE})
\end{eqnarray}
where $Q_{fL}$ and $Q_{fE}$ are the $U(1)_{Y'}$ charges of lepton doublet 
$L$ and lepton singlet $E^{c}$, respectively.
\section{$e^{+}e^{-}\rightarrow h_{k}{\it{l}}^{+}{\it{l}}^{-}$ Cross Section}
We shall analyze the cross section for the $e^{+}e^{-}\rightarrow 
h_{k}{\it{l}}^{+}{\it{l}}^{-}$ scattering, assuming that it proceeds via 
the following chain
\begin{eqnarray}
e^{+}e^{-}&&\rightarrow Z_{i}^{*}\rightarrow h_{k} Z_{j}^{*}\rightarrow 
h_{k}{\it{l}}^{+}{\it{l}}^{-}.\nonumber\\
\end{eqnarray}
The possibility that the vector bosons might come to mass-shell will be 
taken into account by including their total widths to their propagators.
A straightforward calculation yields the following expression 
for the differential cross section 
\begin{eqnarray}
\frac{d\sigma_{k}}{dx}&=&\frac{1}{36(4\pi)^{3}} 
\sqrt{(1+r_{k}-x)^{2}-4r_{k}}\;\;((1-r_{k})^{2}-2x(r_{k}-5)+x^2)\nonumber\\
&&\sum_{ijmn}K_{kij}K_{kmn}
(a^{m}_{e}a^{i}_{e}+v^{m}_{e}v^{i}_{e})(a^{n}_{l}a^{j}_{l}+v^{n}_{l}v^{j}_{l})
\nonumber\\
&&D_{Z_{i}}(s)  D_{Z_{n}}(s)D_{Z_{j}}(x)  D_{Z_{m}}(x)
\end{eqnarray}
where $x=m^2_{l^{+}l^{-}}/s$ is the invariant dilepton mass normalized to $s$,
$r_{k}=m_{h_{k}}^{2}/s$, and 
\begin{eqnarray}
D_{Z_{i}}(s)=\frac{1}{s-M_{Z_{i}}^{2}+iM_{Z_{i}}\Gamma_{Z_{i}}}
\end{eqnarray}
is the propagator of the vector boson $Z_{i}$ and $\Gamma_{Z_{i}}$ is its 
total width. In the cross section differential in (12), $x$ has the range  
$0\le x \le  (1-\sqrt{r_{k}})^{2}$ as follows from the kinematics of the 
decay.

For each value of $k$, using the coupling constants $K_{kij}$ listed in (4),
we can evaluate $\frac{d\sigma_{k}}{dx}$ from (12). In the limit of small 
mixing angle, equations (5) and (6) yield the following vector and 
axial couplings:
\begin{eqnarray}
v^{(1)}_{f}&=&v_{f}\;\; ; \;\; a^{(1)}_{f}=a_{f}\nonumber\\
v^{(2)}_{f}&=&v'_{f}\;\; ; \;\; a^{(2)}_{f}=a'_{f}.
\end{eqnarray} 
As is seen from (9) and (10), vector and axial couplings of $Z'$ boson 
depends on the unknown $U(1)_{Y'}$ charges of the lepton superfields, as 
expected. The family independent $Z'$ models are severely contrained by the 
Z-pole observables. However, if $U(1)_{Y'}$ charges are family- dependent, 
then $Z'$ models are less constrained by the LEP data. Indeed, such 
family dependent extra $U(1)$'s are realizable in string models 
\cite{faraggi,lang}. Alternatively, one can assume leptophobic $Z'$ 
\cite{bar-ber} and neglect these  $U(1)_{Y'}$ lepton charges. Based on 
these reasonings, we shall assume vanishing axial and vector leptonic 
couplings for $Z'$:
\begin{eqnarray}
v^{(2)}_{f}&=&0\;\; ; \;\; a^{(2)}_{f}=0,
\end{eqnarray}
which implies that $\frac{d\sigma_{k}}{dx}$ in (12) gets no contribution 
from $Z_{2}$ boson. Thus, the production of $h_{1}$ and $h_{3}$, with 
the diagonal couplings mentioned before, proceed by the contribution of 
the $Z_{1}$ boson only. The difference between $h_{1}$ and $h_{3}$ comes 
mainly from the dependence of their masses on $h_{S}$ and $\rho_{S}$ 
(See equation (3)). As opposed to $h_{1}$ and $h_{3}$, $h_{2}$ production 
is impossible as its production cross secton identically vanishes due 
to its vanishing couplings to vector bosons in the present minimum of 
the potential.
  
That only $h_{2}$ contributes to pair production process 
$e^{+}e^{-}\rightarrow h A$ \cite{biz}, and no contribution comes to 
Higgs-strahlung $e^{+}e^{-}\rightarrow h_{k}Z_{i}$ from $h_{2}$ is a 
special case of the complementarity between the two processes, as already 
noted in MSSM \cite{mssm-strah}. In the scattering process under concern, 
a non-zero $h_{2}$ signal would show up when the lepton superfields attain 
non-zero $U(1)_{Y'}$ charges and mixing angle is large enough. However, 
when these two conditions are satisfied one expects non-negligable shifts in 
the Z-pole observables in such a light $Z'$ model \cite{kolda}. 

In analyzing $e^{+}e^{-}\rightarrow h_{k}{\it{l}}^{+}{\it{l}}^{-}$ 
scattering process we first integrate the differential cross 
section given in (12) over the entire range of $x$ to obtain the total 
cross section. Next we depict the dependence of the total cross section 
$\sigma$ for each Higgs on $h_{S}$ and $\rho_{S}$. In this way we can 
investigate the dependence of the cross section on $\mu$ parameter and $Z'$ 
mass. In the numerical analysis we take a lower bound of $60\; GeV$ for 
$m_{h_{1}}$ which implies a lower bound of $h_{S_{min}}\sim 0.345$ via (3). 

Depicted in Fig. 1 is the $h_{S}$ dependence of the $h_{1}$ production cross 
section, $\sigma_{h_{1}}$ for $\sqrt{s}=130$, 136, 161, 172, 183, 205 $GeV$ 
from left to right. The first five points have been probed at LEP2 and 
the last one is planned to be under search soon. In forming Fig. 1 we 
let $h_{S}$ vary from 0.345 (corresponding to $m_{h_{1}}\sim 60\;GeV$) 
up to the kinematical threshold $\sqrt{sG^2}/(2 M_{Z})$ (corresponding to
$m_{h_{1}}\sim \sqrt{s}$). For each $\sqrt{s}$ value the cross section 
vanishes at some value of $h_{S}$ at which the Higgs production stops. 
As is seen from the figure, for $\sqrt{s}=205\; GeV$ the cross section 
remains around $10\; fb$ up to $h_{S}\sim 0.7$, so that $m_{h_{1}}$ values 
up to $130\; GeV$ can be probed at this CM energy depending on the total 
integrated limunosity of the collider under concern. Similar behaviour 
occurs for $\sqrt{s}=161$, 172, and 183 $GeV$ for $h_{S}$ less than 0.4, 
0.45 and 0.55, respectively. 
  
Being much heavier than $h_{1}$, production of $h_{3}$ is very hard at these 
CM energies. To demonstrate the behaviour of the cross section we 
plot $\sigma_{h_{3}}$ as a function of $h_{s}$ for $\sqrt{s}=183\; GeV$ 
and various values of $\rho_{s}$. Using $r_{3}\le 1$,
$M_{Z_{2}}\ge M_{Z}$, and $h_{S}\ge 0.345$, it is easy to find the
ranges of $h_{S}$ and $\rho_{S}$: $0.58\le \rho_{S}\le 1.17$ and $0.87\ge 
h_{S} \ge 0.345$. As is seen from the figure, the cross section is much 
smaller than $\sigma_{h_{1}}$, and vanishes for certain $h_{S}$ values as 
in Fig. 1. For small $h_{S}$ the largest cross section occurs for 
$\rho_{S}=0.58$ at $h_{S}=0.345$. For higher values of $\rho_{S}$, cross 
section gets maximized at the higher end of the $h_{S}$ range. It 
is clear that with such a small cross section, it is necessary to have a 
large limunosity to obtain sufficient number of events differing 
hopefully from the predictions of the SM. 
 
As is generally the case, unless one resorts to some sophisticated numerical 
techniques (like the PYHTIA and JETSET packages), the theoretical results  
obtained in this way can hardly be confronted with the experiment directly, 
because of the various cuts applied in event selection process to 
suppress the unwanted background. Indeed, as one of the latest LEP2 
releases \cite{OPAL} shows, the Bjorken process under concern has a large 
SM background and Higgs search in MSSM and 2HDM for $\sqrt{s}=130-172\; GeV$ 
has a negative result. However, the results of the present work 
may serve as a guide for the design of near-future colliders like NLC.

In the above we have mainly focused on unpolarized electron- 
positron beams in which cross section differential $d\sigma/d\Omega$ 
involves only one angular variable: the angle between Higgs particle and 
the electron beam. When the initial beams are trasversely polarized, 
however, $d\sigma/d\Omega$ picks up an explicit dependence on the Higgs 
boson azimuthal angle. Therefore, in this case, in addition to the cross 
section itself, one can work out the azimuthal asymmetry of the process, 
which provides further information on the appropriate portion of the 
parameter space of the model under concern. However, let us note that the 
total cross section for transversely polarized beams is the same as the 
one for unpolarized beams. If the electron-positron beams are logitudinally 
polarized, angular dependence of $d\sigma/d\Omega$ is the same as in the 
unpolarized case. However, the difference between the two beam 
polarizations enables one to define spin asymmetry of the process which 
can be important in constraining the parameter space of the model. Unlike 
the case of the transverse polarization of the beams, here the total cross 
section  differs from that of the unpolarized case. The SM- counterpart 
of the process under concern was analyzed by \cite{guliev} with a detailed 
discussion of the polarization effects and it seems unnecessary to 
have a detailed discussion of the polarization effects here because 
results of \cite{guliev} applies equally to the $Z^{\prime}$ model 
under concern.
     
Until now our analysis has been based entirely on the tree-level 
potential in the limit of large trilinear Higgs coupling, which has 
already been shown to follow from string-inspired models in 
\cite{lang} through the solution of RGE's. However, it is known that the 
results obtained using the tree-level potential are very sensitive to 
the choice of the renormalization scale $Q$ of the RGE's, and hence one 
should take into account the effects of the loop corrections on the 
scalar potential to get more reliable results. This is what we will 
do next, up to one-loop order, in the frame work of the effective 
potential formalism\cite{carena}.

In the framework of the effective potential formalism, the contribution 
of the radiative corrections to the scalar potential is given by 
\begin{eqnarray}
\Delta V = \frac{1}{64\pi^2}
Str{\cal{M}}^{4}[\ln\frac{{\cal{M}}^{2}}{Q^2}-C]\; ,
\end{eqnarray}
where $Str$ means the usual supertrace, ${\cal{M}}^{2}$ is the field  
dependent mass-squared matrix of the fields, and $C$ is a constant 
that depends on the renormalization scheme \cite{carena}. The dominant 
contribution to the effective potential comes from the top quark and stops, 
due to relatively large value of the top Yukawa coupling. In the limit of 
degenerate stops, the most significant one-loop corrections arise for 
Higgs trilinear coupling $A_{s}$, $H_{2}$ quartic coupling 
$\lambda_{2}$, and $H_{2}$ soft mass-sqared $m_{2}^{2}$, namely
\begin{eqnarray}
\hat{A}_{s}&=&A_{s}+\beta_{h_{t}}S_{t\tilde{t}}A_{t}\nonumber\\
\hat{m}_{2}^{2}&=&m_{2}^{2}+\beta_{h_{t}}[
(A^{2}+A_{t}^{2})S_{t\tilde{t}}-A^{2}]\\
\hat{\lambda}_{2}&=&\lambda_{2}+\beta_{h_{t}}S_{t\tilde{t}}h_{t}^{2}\nonumber
\end{eqnarray}
where $\beta_{h_{t}}=\frac{3}{(4\pi)^{2}}h_{t}^{2}$, $S_{t\tilde{t}}=  
\ln\frac{m_{\tilde{t}_{1}}m_{\tilde{t}_{2}}}{m_{t}^{2}}$ is the top-stop 
splitting function, and  $A^{2}=m_{Q}^{2}+m_{U}^{2}$, $m_{Q}^{2}$ 
and $m_{U}^{2}$ being the soft mass-squareds of squark doublet 
$\tilde{Q}$ and  up-type  $SU(2)$ squark singlet $U^{c}$, respectively. 
As was already discussed in \cite{lang}, one can realize the large Higgs 
trilinear coupling minimum at low energies by integrating RGE's with  
non-universal boundary conditions from the string scale down to the weak 
scale. The resulting parameter space satisfies $m_{Q}^{2}\sim 
m_{U}^{2}\sim (h_{t}A_{t})^{2}\sim (h_{s}A_{s})^{2}$ that are much 
larger than the Higgs soft mass-squareds. For this parameter set, it is 
easily seen that $\tilde{m}_{t_{1,2}}^{2}\sim m_{Q}^{2}+m_{t}^{2}$, proving 
the consistency of assuming degenerate stops. Consequently 
$S_{t\tilde{t}}\sim \ln (1+ m_{Q}^{2}/m_{t}^{2})$, and it remains close to  
unity as the squark soft masses are around the weak scale. 

In the light of these results, one can now discuss the stability of the 
large trilinear coupling minimum under the contributions of the radiative 
corrections in (16). From the first equality in (16) one 
observes that $A_{s}$ increases in proportion with $A_{t}$. Similarly, 
from the second equation in (16) one infers that $m_{2}^{2}$ increases 
with $A_{t}^{2}$. Letting $m^{2}=m_{1}^{2}+m_{2}^{2}+m_{S}^{2}$ be the 
sum of the soft mass-squareds of the Higgs fields, and $\lambda$ be the 
sum of the quartic coefficients in the potential, one can show that 
the large trilinear coupling minimum occurs after $A_{s}$ exceeds the 
critical point 
\begin{eqnarray}
A_{s}^{crit}=\sqrt{\frac{8}{9}\frac{\lambda m^2}{h_{s}^{2}}}.
\end{eqnarray} 
This formula is valid when $m^{2}>0$, which is easy to realize at low 
energies if non-universality is permitted at the string level 
\cite{lang}. Actually, such a condition is necessary for the need to an 
$A_{s}$ dominated minimum to arise, as otherwise one can create an 
appropriate low energy minimum by using merely the Higgs soft 
mass-squareds. At the tree-level $\lambda=3h_{s}^{2}$, and at the loop- 
level it increases by $\beta_{h_{t}}S_{t\tilde{t}}h_{t}^{2}$ due to 
the contribution of $\lambda_{2}$ in (16). Similarly, due to the 
contribution of $\hat{m}_{2}^{2}$ in (16), $m^{2}$ increases by 
$\beta_{h_{t}}A_{t}^{2}S_{t\tilde{t}}$. Inserting these loop 
contributions to (17), and using the tree-level relation $A_{s}\sim 
A_{t}$ required by the solution of the RGE's, one gets the following 
expression for the radiatively corrected $A_{s}^{crit}$: 
\begin{eqnarray}
\hat{A}_{s}^{crit}=\frac{A_{s}^{crit}}{1-\frac{1}{3}\beta_{h_{t}} 
S_{t\tilde{t}}}.
\end{eqnarray}
Therefore radiative corrections magnify the critical point that $A_{s}$ 
must exceed to develop a large trilinear coupling minimum. If the 
tree-level value of $A_{s}$ is close to the critical point, radiative 
corrections can destabilize the vacuum. For example, for 
$m_{1}^{2}=m_{2}^{2}=m_{s}^{2}=m_{0}^{2} >0$ one gets $A_{s}^{crit}\sim 
2.82\, m_{0}$. While for $A_{s}<A_{s}^{crit}$, all VEV's identically vanish, 
for $A_{s}>A_{s}^{crit}$ all VEV's are equal and get the value of 
$A_{s}/(h_{s}\sqrt{2})$. Even if $A_{s}>A_{s}^{crit}$, when it is 
in the vicinity of $A_{s}^{crit}$, radiative corrections cause a 
transition from the broken phase to the symmetric phase, and completely 
restores the symmetry. However, for large enough $A_{s}$, there occurs no 
problem in the implementation of the minimum of the potential under 
concern. Thus, by choosing $A_{s}$ large enough one can analyze the 
physical processes with reliable accuracy using the tree-level potential.

Recently, in \cite{ma}, electroweak breaking in E(6)- based $Z'$ 
models together with radiative corrections to scalar masses was 
discussed. Here we consider a general string-inspired family-dependent 
$Z'$ model without exotics \cite{lang} in which we do not restrict 
ourselves to E(6) charge assignments. One notes that the 
minimum value of $h_{S}$, $h_{S}^{min}=0.345$, which we have used in forming 
the graphs turns out to be the 'typical' value for $h_{S}$ in such models.  

In all extensions of the SM namely,  2HDM, MSSM, NMSSM, and $Z^{\prime}$ 
models,  $h Z Z$ and $h A Z$ couplings are smaller than those of the SM 
due to  the mixings in the Higgs sector (and also in the vector boson 
sector in  $Z^{\prime}$ models) of the model, and this cause a reduction 
in the associated cross sections compared to those in the SM.  For 
example, in MSSM \cite{giudice}, lightest Higgs production cross 
section is suppressed by the exitence of the factor  
$\sin^{2}(\beta-\alpha)$, where $\alpha$ is the CP-even Higgs mixing 
angle. In addition to the mixing-induced reduction in the cross sections, 
one has a multitude of scalar particles which make it hard to 
differentiate between the underlying models. 

Let us first summarize the situation in the $Z^{\prime}$ models and then 
discuss the other models briefly. As was discussed in \cite{biz}, the 
Higgs pair production process allows only $h_{2}$ production in the 
limit of large Higgs trilinear coupling. Unlike this, however, the 
Bjorken process discussed here forbids the $h_{2}$ production, which 
make two processes complemetary to each other. Hence in the Higgs- 
strahlung type processes, as the collider energy sweeps a certain range 
of values, one can get at most two CP- even scalars at the final state.  

In MSSM, there does not exist a minimum of the potential like the 
one discussed here. However, in the limit of large Higgs bilinear 
mass term, $\mu B$ (decoupling limit), there is a single light scalar in 
the spectrum, all other scalars being much heavier. In this sense, MSSM 
mimics the SM at low energies concerning their Higgs sectors. In the 
opposite limit, one has a  relatively light Higgs spectrum consisting of 
two CP-even and one CP- odd scalar, whose existence are to be probed at the 
colliders. In the absence of similarities between $Z^{\prime}$ models and 
MSSM in the large Higgs trilinear coupling minimum of the former, the 
only observable difference between the two models relies on the number of 
scalars they predict. 
 
In NMSSM, Higgs sector is larger than that of the MSSM, due to the 
existence of the SM-singlet superfield $S$. Indeed, NMSSM has three CP-even, 
and two CP-odd scalars which make it completely different than MSSM and 
$Z^{\prime}$ models (due to one extra CP- odd scalar). However, in 
discussing the Higgs-strahlung process (where CP-odd Higgs effects are 
absent) NMSSM and $Z^{\prime}$ models are expected to behave in a similar 
manner as they have the same number of CP-even scalars. Given the 
experimental signature of the Higgs-strahlung process under concern over 
a certain energy range, is it possible to differentiate between NMSSM and 
$Z^{\prime}$ models ? In the limit of large Higgs trilinear coupling, it 
is possible to have a parallel analysis of these two models from which one 
may conclude on the dissimilarities between the two. It is clear that to 
have a proper large Higgs trilinear coupling minimum, in which all scalars 
have the VEV's of similar magnitude, it is necessary to suppress the 
contribution of $S^{3}$ term \cite{nmssm-strah} to the soft supersymmetry 
breaking part of the scalar potential. With this restriction in mind and 
assuming a small but non-vanishing Yukawa coupling, $k$, for the $S^{3}$ 
term (to forbid the creation of an axion) in the superpotential, one finds 
the following $h Z Z$ couplings, 
\begin{eqnarray}
K_{111}=\frac{G M_{Z}}{2}\;,\;\; K_{211} = 0\;, \;\;K_{311}=0
\end{eqnarray}
corresponding to the first three couplings in (4). The mass spectrum now is 
given by 
\begin{eqnarray}
m_{h_{1}}\sim\frac{\sqrt{2}h_{S}M_{Z}}{G}\; , \;
m_{h_{2}}\sim\frac{\sqrt{2 h_{S}^{2}+4k^{2}}}{G}M_{Z}\; , \;
m_{h_{3}}\sim\frac{\sqrt{G^2+2 h^{2}_{S}}}{G}M_{Z},
\end{eqnarray}
so that the lightest Higgs has the same mass in both models. However, as 
dictated by (20), in the NMSSM only the lightest Higgs  contributes to
the Higgs- strahlung process under discussion. This forms a spectacular 
difference between the two models: while one expects two CP-even Higgs 
particles at the final state in $Z^{\prime}$ models, in NMSSM there is 
just one Higgs particle. 

Despite the negative Higgs search at LEP2, the Bjorken process discussed 
here will continue to be one of the main Higgs search tools in future 
colliders.  Therefore it seems necessary to have the predictions of  
various models in hand for comparison with the near-future collider data. 
In conclusion, we have analyzed Higgs production through the Higgs-strahlung 
type processes in $Z^{\prime}$ models in that minimum of the potential 
for which Higgs trilinear coupling is large  compared to the other soft 
mass parameters. At the same type of minimum, NMSSM is found to have 
observable diffrences with the $Z^{\prime}$ models.

\newpage
\begin{figure}
\vspace{5cm}
\end{figure}
\begin{figure}
\vspace{12.0cm}
    \includegraphics{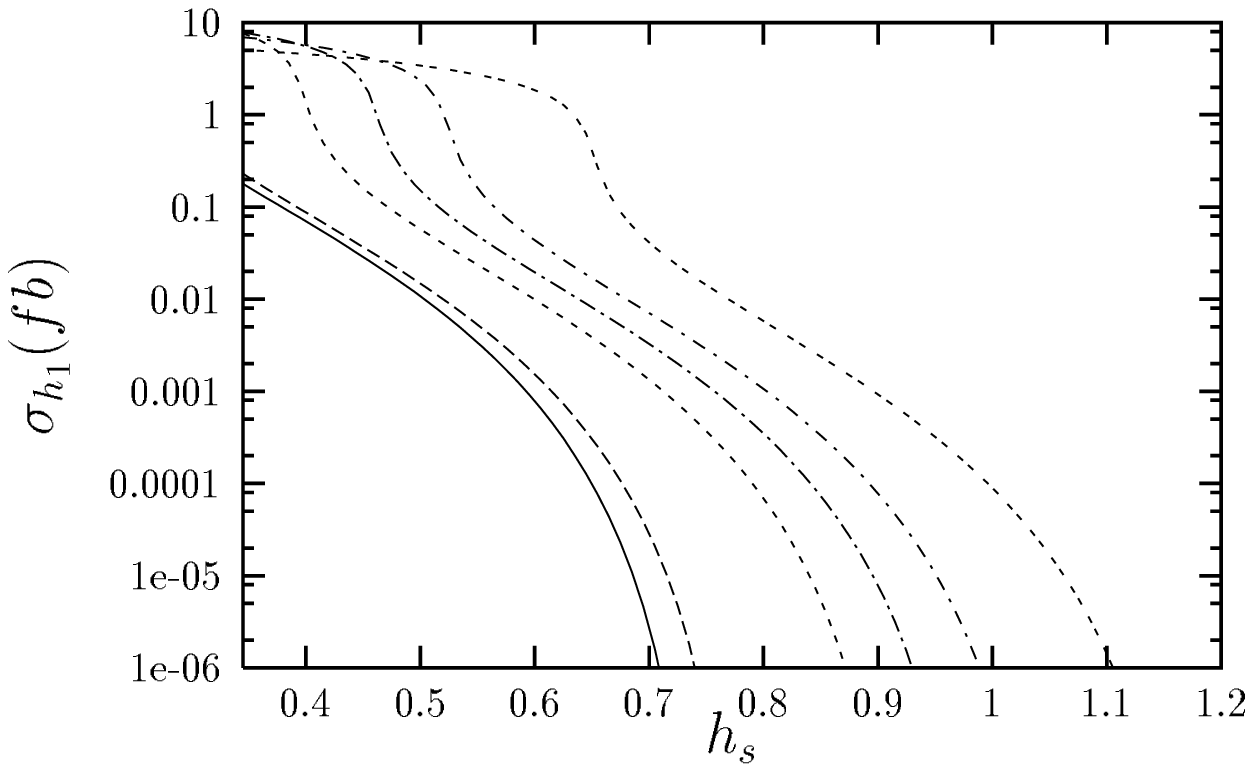}  
    \vspace{-8.0cm}
\vspace{0.0cm}
\mbox{ \hspace{0.5cm} \large{\bf Figure 1: Dependence of $\sigma_{h_{1}}$ on
$h_{S}$ for $\sqrt{s}=$130, 136,}} 
\mbox{ \hspace{0.5cm} \large{\bf 161, 172, 183 and 205 $GeV$ from left to 
right.}} 
\end{figure}
\begin{figure}
\vspace{12cm}
\end{figure}
\begin{figure}
\vspace{12.0cm}
    \includegraphics{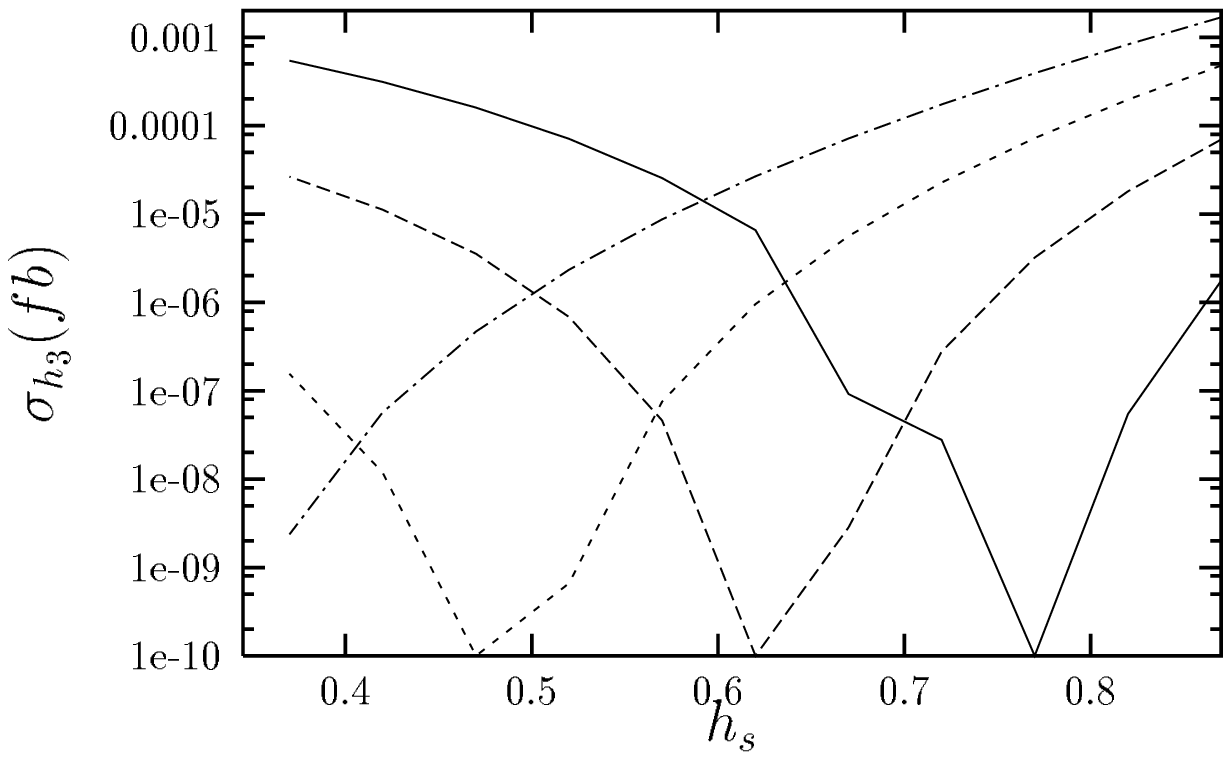}
    \vspace{-8.0cm}
\vspace{0.0cm}
\mbox{ \hspace{0.5cm} \large{\bf Figure 2: Dependence of 
$\sigma_{h_{3}}$ on $h_{S}$ for $\sqrt{s}=183\;GeV$}}
\mbox{ \hspace{0.5cm} \large{\bf and 
 $\rho_{S}=$ $0.58$ (solid), $0.8$ (dashed), $1.0$ (dotted)}} 
\mbox{ \hspace{0.5cm} \large{\bf
and $1.17$ (dot-dashed).}} 
\end{figure}
\end{document}